%% file: paper.tex
\documentclass[sigconf,screen,nonacm]{acmart}

\usepackage{graphicx}
\usepackage{booktabs}
\usepackage{tabularx}
\usepackage{makecell}
\usepackage{pifont}
\usepackage{threeparttable}
\usepackage[referable]{threeparttablex}

\newcolumntype{L}{>{\raggedright\arraybackslash}X}
\newcolumntype{C}{>{\centering\arraybackslash}X}
\usepackage{xspace}
\usepackage{xcolor}
\usepackage{colortbl}
\usepackage{longtable}
\hypersetup{colorlinks=true,linkcolor=black,urlcolor=blue,citecolor=black}
\usepackage{wasysym}

\newcommand{\parag}[1]{\vspace{2mm}\noindent\textbf{#1.}\hspace{2mm}}
\newcommand{\SCI}{\texttt{SCI}\xspace}
\newcommand{\RSCI}{\texttt{rSCI}\xspace}
\newcommand{\OSCI}{\texttt{oSCI}\xspace}

\newcommand{\E}{\texttt{E}\xspace}
\newcommand{\I}{\texttt{I}\xspace}
\newcommand{\M}{\texttt{M}\xspace}
\newcommand{\R}{\texttt{R}\xspace}
\newcommand{\faithful}{\textcolor[HTML]{1a7f37}{\ding{51}}}
\newcommand{\conditional}{\textcolor[HTML]{1a7f37}{\ensuremath{\circ}}}
\newcommand{\proxyonly}{\ensuremath{\sim}}
\newcommand{\notrefl}{\textcolor[HTML]{d1242f}{\ding{55}}}
\newcommand{\undocumented}{\textcolor[HTML]{555555}{\textbf{?}}}
\newsavebox{\cfabox}
\newenvironment{callforaction}
  {\par\medskip\noindent\begin{lrbox}{\cfabox}%
   \begin{minipage}{\dimexpr\columnwidth-2\fboxsep-2\fboxrule\relax}%
   \small\textbf{Ask.}\hspace{1mm}\itshape}
  {\end{minipage}\end{lrbox}\noindent\fbox{\usebox{\cfabox}}\par\medskip}

\AtBeginDocument{%
  }

\setcopyright{none}
\fancypagestyle{arxivfirstpage}{%
  \fancyhf{}%
  \fancyfoot[L]{\raisebox{-6pt}[0pt][0pt]{\parbox[t]{\columnwidth}{\footnotesize Extended version of our vision paper accepted for publication at the\\ACM Symposium on Cloud Computing (SoCC '26). DOI: \href{https://doi.org/10.1145/3828161.3857847}{10.1145/3828161.3857847}}}}%
  \fancyfoot[C]{\footnotesize\thepage}%
}

\begin{document}
\title{Reconciling Bottom-Up Metrics with Top-Down Reporting for Cloud Carbon Accounting}

\author{Philipp Wiesner}
\affiliation{%
  \institution{University of Cambridge}
}
\affiliation{%
  \institution{Logsight.ai}
}

\author{Loïc Lannelongue}
\affiliation{%
  \institution{University of Cambridge}
}

\author{Alexander Acker}
\affiliation{%
  \institution{Logsight.ai}
}

\author{Odej Kao}
\affiliation{%
  \institution{TU Berlin}
}

\begin{abstract}
Organizations seeking to reduce the carbon footprint of their cloud applications must rely on two largely disconnected ways of measuring it, each with important limitations.
Bottom-up metrics like Software Carbon Intensity (\SCI) provide signals for carbon-aware optimization, but tenants lack the information needed to account for many provider-side overheads.
Top-down reports by cloud providers provide a more comprehensive view of a tenant's carbon footprint but are coarse and methodologically opaque.
Because the two approaches differ in scope and reporting frequency, the current state of the art treats them as decoupled: bottom-up metrics for optimization, top-down reports for corporate disclosure.

We argue that the two signals should be \emph{reconcilable}.
A per-work\-load metric whose improvements never surface in the provider's audited report will not be adopted for accountability at scale.
We survey the state of the art in cloud carbon accounting and propose a vision for reconciled \SCI (\RSCI): a per-workload metric that re-anchors a bottom-up energy estimate to the cloud provider's top-down report through a residual.
By decomposing and allocating this residual according to its physical drivers, \RSCI can preserve operational incentives while also attributing idle capacity and embodied carbon to the workloads that drive them.
We show that today's top-down reporting is still too coarse and methodologically inconsistent to support credible reconciliation and lay out concrete steps in provider reporting and standards to make it practical.
\end{abstract}

\maketitle
\thispagestyle{arxivfirstpage}

\section{Introduction}
\label{sec:intro}

A major challenge in sustainable computing is quantifying the carbon footprint of workloads.
It is a prerequisite for determining the effectiveness of carbon-reduction strategies, for reporting to regulators and investors, and for making informed decisions about workload scheduling and adaptation.
Two largely separate approaches exist: bottom-up modeling and top-down corporate greenhouse gas (GHG) reporting.

\begin{figure}[t]
  \centering
  \includegraphics[width=.95\columnwidth]{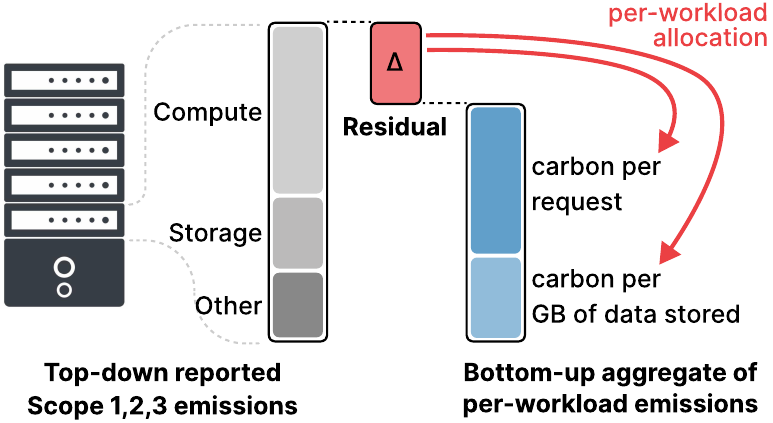}
\vspace{-1mm}
  \caption{A provider's top-down carbon report and a tenant's bottom-up estimate rarely match. We call for \emph{reconciling} the two totals and allocating the residual back to workloads.}
\vspace{-3mm}
\end{figure}

\parag{Bottom-up metrics}
Most carbon-aware computing research has focused on optimizing bottom-up metrics like operational carbon~\cite{radovanovic2022google,hanafy2024gaia,sukprasert2024limits,gsteiger2024caribou,murillo2024cdn,lechowicz2023pause,hanafy2024carbonscaler,wiesner2025qualitytime,wiesner2025moving}: the estimated emissions from a workload's marginal increase in energy consumption.
This signal is straightforward to obtain but does not capture overheads from idle hardware or embodied emissions.
To provide a more comprehensive metric, the Green Software Foundation proposed the Software Carbon Intensity (\SCI) specification~\cite{sci-spec}, codified as ISO/IEC~21031 in 2024, which expresses the carbon emitted per unit of work and folds in an embodied term.
However, Bashir et al.~\cite{bashir2024} show that \SCI can produce perverse incentives for workload placement and scaling in practice.
Furthermore, every bottom-up metric, \SCI included, can only capture what a tenant can observe or model.

\parag{Top-down reporting}
Due to regulatory pressure, cloud providers increasingly offer customer-facing dashboards and reports enabling tenants to estimate and disclose their cloud carbon footprint in line with the GHG Protocol~\cite{ghg-protocol-corp}.
While the reporting boundaries and allocation methods can differ significantly by provider, the general methodology is usually public and sometimes even externally audited.
However, the reported numbers are highly aggregated:
reports mostly contain a single number per month, region, customer, and GHG Protocol scope, sometimes coarsely categorized further into service categories.
The reports are typically published about 14-21 days after month-end, which makes them comprehensive but unsuitable for any type of real-time optimization.

\parag{The need for reconcilability}
Today, when an engineering team improves the bottom-up metric of an internal workload, it is entirely unclear whether that improvement reaches the provider's top-down report.
The two can diverge in both directions.
The top-down report contains overheads that bottom-up models cannot cover without internal provider data.
And conversely, the actions a bottom-up metric does capture, such as lowering a VM's utilization and thus its power draw, may not move the reported number at all, if the provider allocates carbon to tenants by usage-hours or revenue instead of measured energy.

The Green Software Foundation explicitly endorses this separation, distinguishing ``measuring for action'' (per-workload signals that drive engineering) from ``measuring for reporting'' (corporate carbon accounting) and emphasizing that the two serve different goals and should not be conflated~\cite{gsf-measuring}.
\emph{We argue that different purposes should not mean disconnected accounting.}
Regulatory disclosure requirements~\cite{eu2022csrd} create a need for workload-level signals that attribute reported cloud emissions across internal teams and products.
A per-workload metric whose gains never land in official reporting will not be trusted and adopted for accountability at scale.

\parag{Contributions}
We argue that bottom-up carbon metrics should \emph{reconcile} with the provider's top-down GHG report by construction.
Towards this, we make the following contributions:
\begin{itemize}
\item We review the state of the art in cloud carbon accounting, including a survey of 50 providers and a matrix showing which engineering actions are reflected in providers' documented accounting methods.
\item We introduce \emph{reconciled Software Carbon Intensity} (\RSCI), which anchors a bottom-up energy estimate to the provider's top-down report through a \emph{residual}, so per-workload emissions sum to the reported total by construction.
\item We decompose the residual into energy-driven and capacity-driven components and propose allocating them according to their drivers, so \RSCI can account not only for the energy a workload consumes, but also for the idle capacity and embodied hardware the workload contributes to.
\item We diagnose why today's provider reports are too coarse and methodologically opaque for computing credible \RSCI scores in practice, and set out concrete asks on reporting and standards.
\end{itemize}

 \section{Bottom-Up Metrics Miss Provider Overheads}
\label{sec:background}

A bottom-up metric estimates a workload's carbon from quantities a tenant can measure without access to providers' internal data.
The de facto choice in most research papers is operational carbon, which we write as \OSCI (operational software carbon intensity) to match the terminology in Bashir et al.~\cite{bashir2024}:
\[
\OSCI \;=\; \frac{\E \cdot \I}{\R},
\]
the workload's energy $\E$ times its region's carbon intensity $\I$ (in gCO$_2$/kWh), per functional unit of work $\R$ (an API call, a training run, a GB stored per month).\footnote{We write CO$_2$ or ``carbon'' as shorthand for CO$_2$-equivalent emissions throughout.}
It is straightforward, transparent, and it responds to the two operational dimensions a tenant controls directly: energy efficiency and carbon-aware timing or placement.
But it says nothing about hardware efficiency, i.e., whether the workload requires less provisioned capacity.

To capture this missing dimension, the Software Carbon Intensity (\SCI) specification (ISO/IEC~21031~\cite{sci-spec}) adds an embodied term:
\[
\SCI \;=\; \frac{\E \cdot \I \;+\; \M}{\R},
\]
where $\M$ is the workload's amortized share of the hardware's embodied carbon, allocated according to its reserved time and resource share as described by Gupta et al.~\cite{gupta2022act}.
The specification defines three categories of \emph{action} for reducing software carbon: (i) energy efficiency, (ii) hardware efficiency, and (iii) carbon awareness, and requires the score to fall whenever a tenant takes any of these actions~\cite{sci-spec}.
\OSCI already responds to the first and third through \E and \I, the embodied term \M is what extends \SCI to the second.

However, Bashir et al.~\cite{bashir2024} identify a \emph{sunk-carbon fallacy} when past embodied emissions guide operational scheduling.
A lower amortized embodied term can favor a less energy-efficient server, even though scheduling cannot undo its manufacturing emissions.
They therefore recommend sticking to \OSCI for operational decisions such as scheduling and scaling.
But \OSCI is incomplete: it rewards neither reserving less hardware (beyond the energy it saves), nor purchasing less and lower-embodied hardware.

Most importantly, \emph{bottom-up estimates do not guarantee complete attribution of provider emissions}.
Tools and services such as Cloud Carbon Footprint~\cite{cloudcarbonfootprint}, BoaviztAPI~\cite{simon2024boaviztapi}, and Greenpixie~\cite{greenpixie-cloud-methodology} help tenants estimate operational and embodied emissions, but accounting for shared idle capacity and other provider-side overheads requires information that tenants cannot directly observe.
Even operational estimates can disagree with provider allocations: Jacquet et al.~\cite{jacquet2025cinergy} estimate up to three times as much carbon for busy VMs as OVHcloud's method, which allocates emissions by VM size.
Adding an embodied term does not resolve these gaps or establish consistency with the provider's inventory.

\section{Top-Down Reports Respond Inconsistently to Carbon-Reduction Actions}
\label{sec:sota}

To support customer emissions disclosure, cloud providers have moved beyond annual corporate reports toward customer-facing tools that attribute emissions to individual tenants.
But a top-down report is only useful for engineering if it responds to the same actions a per-workload metric rewards.
We surveyed 50 cloud providers in September~2026, spanning general-purpose compute, GPU/HPC, CDN/edge, application hosting, and object storage.
Interestingly, none of the 10 GPU/HPC (neocloud) providers we surveyed expose per-tenant emissions.
Appendix~\ref{app:inventory} gives the full provider list, inclusion criteria, and source evidence.
Table~\ref{tab:levers} compares the eleven surveyed providers with customer carbon reporting for general-purpose compute, showing which actions their methods reflect.
Appendix~\ref{app:action-evidence} explains the classification.

We find that no provider responds to all actions, and the same action can be reflected inconsistently across providers’ reports.
This is because a top-down report is built in three steps: providers decide \emph{what} emissions to count, \emph{allocate} the datacenter total down to services and tenants, and \emph{report} the result at a chosen granularity.
In practice, methodologies diverge at each step.

\parag{What cloud providers cover}
Providers decompose their total datacenter emissions over the reporting period along the GHG Protocol's three scopes.
\emph{Scope~1}, direct on-site emissions such as backup-diesel combustion and refrigerant leakage, is usually negligible for most tenants today, though the increasing deployment of behind-the-meter gas generation at datacenters may change this in future.
\emph{Scope~2} covers emissions from purchased electricity.\footnote{The GHG Protocol mandates dual reporting for Scope~2: \emph{location-based} figures reflect actual grid emissions, while \emph{market-based} figures net out procurement instruments like power purchase agreements (PPAs) and renewable energy certificates (RECs), which in practice often moves the market-based number to near zero. Since engineering teams have no control over the provider's procurement choices, the market-based figure is not actionable for workload optimization, which is why \SCI explicitly excludes it~\cite{sci-spec}. We only consider the location-based figure in this paper.}
\emph{Scope~3} covers amortized cradle-to-gate hardware and, depending on the provider, building embodied carbon, upstream fuel-and-transmission losses (FERA), freight, and end-of-life.
No two of the three hyperscalers count the same set of Scope-3 categories.
For example, only GCP counts employee business travel and commuting~\cite{google-carbon-methodology}, only Azure counts hardware end-of-life~\cite{microsoft-s3-2021}, and building-embodied carbon is counted by AWS and GCP but not Azure~\cite{aws-ccft-model3,google-carbon-methodology}.
Appendix~\ref{app:reporting-coverage} compares the documented emissions coverage, electricity accounting methods, and reporting detail across the surveyed providers offering customer reports.

\parag{How cloud providers allocate}
Providers then allocate the datacenter total to tenants in proportion to their share of an \emph{activity unit}.
The units are inconsistent and lean on economics to differing degrees.
For example, AWS allocates foundational services by usage units, including normalized instance-hours for EC2, and non-foundational services by ``equivalent revenue''~\cite{aws-ccft-model3}.
GCP distributes each product's energy across SKUs in proportion to list price before allocating by per-SKU usage such as vCPU-hours~\cite{schneider-mattia-2024}.
This choice is at least transparent and predictable: a tenant can reliably reduce its allocation by shifting to cheaper SKUs, even if it usually cannot tell which SKU is actually more energy-efficient.

\begin{table}[t]
\centering
\caption{The \SCI defines three \emph{action} categories for reducing carbon.
The table shows whether providers' documented accounting methods reflect these actions.
\faithful~supported, \proxyonly~partial or proxy response, \notrefl~unsupported, \undocumented~undocumented.}
\vspace{-3mm}
\label{tab:levers}
\small
\setlength{\tabcolsep}{3.5pt}
\renewcommand{\arraystretch}{1.2}
\begin{tabular*}{\columnwidth}{@{\extracolsep{\fill}}l cc c cc@{}}
\toprule
 & \multicolumn{2}{c}{\makecell{\textbf{Energy}\\\textbf{efficiency}}} & \makecell{\textbf{Hardware}\\\textbf{efficiency}} & \multicolumn{2}{c}{\makecell{\textbf{Carbon}\\\textbf{awareness}}} \\
\cmidrule(lr){2-3}\cmidrule(lr){4-4}\cmidrule(lr){5-6}
\textbf{Provider} & \makecell{Efficient\\resource} & \makecell{Lower\\utilization} & \makecell{Right-\\size} & Region & Time \\
\midrule
AWS~\cite{aws-ccft-model3}      & \proxyonly & \notrefl  & \faithful & \faithful & \notrefl \\
GCP~\cite{google-carbon-methodology}      & \proxyonly   & \notrefl  & \faithful & \faithful & \notrefl \\
Azure~\cite{azure-methodology-2026,microsoft-s3-2021} & \proxyonly & \notrefl & \faithful & \notrefl & \notrefl \\
Alibaba~\cite{alibaba-cloud-carbon-footprint} & \proxyonly & \undocumented & \undocumented & \notrefl & \notrefl \\
IBM~\cite{ibm-carbon-calc-v3}      & \proxyonly & \notrefl  & \faithful & \faithful & \notrefl \\
Oracle~\cite{oracle-emissions-management} & \proxyonly & \undocumented & \proxyonly & \faithful & \undocumented \\
\midrule
OVH~\cite{ovh-eit-v2} & \proxyonly & \notrefl & \faithful & \faithful & \notrefl \\
Scaleway~\cite{scaleway-instances-footprint,scaleway-calculation-values} & \proxyonly & \proxyonly & \faithful & \faithful & \notrefl \\
T~Cloud~\cite{tcloud-sustainability-dashboard,tcloud-sustainability-release} & \proxyonly & \undocumented & \proxyonly & \notrefl & \notrefl \\
Exoscale~\cite{exoscale-impact-methodology} & \proxyonly & \notrefl & \faithful & \faithful & \notrefl \\
OUTSCALE~\cite{outscale-carbon-methodology} & \undocumented & \undocumented & \undocumented & \faithful & \notrefl \\
\bottomrule
\end{tabular*}
\vspace{-2mm}
\end{table}

Where VM allocation depends only on reserved resources or cost, two VMs of the same type and duration receive the same allocation weight, even if one is idle and the other fully loaded.
Oracle describes allocating measured hardware energy to workloads without specifying the customer weights~\cite{oracle-emissions-management}.
Scaleway models CPU-instance power from utilization~\cite{scaleway-instances-footprint}, while T~Cloud uses service models calibrated from infrastructure measurements~\cite{tcloud-sustainability-release,tcloud-sustainability-dashboard}.
Scaleway's bare-metal methodology~\cite{scaleway-baremetal-footprint} is a clean example: because the customer controls a dedicated physical server, Scaleway can attribute that server’s operational energy and amortized manufacturing footprint directly to the customer.

The activity unit also decides the \emph{efficient-resource} action: switching to a more efficient instance lowers the number only where the per-unit factor is grounded in energy or LCA, whereas at GCP it tracks list price, so a cheaper-but-dirtier SKU can look greener~\cite{schneider-mattia-2024}.

\parag{How cloud providers report}
Most customer dashboards report at the service level (for GCP, ``Compute Engine'', ``Networking'', ...), far coarser than a bottom-up metric's per-workload view.
A tenant sees Scope 1, 2, and 3 totals per service $\times$ region $\times$ month, released 14 to 21 days after the period closes.
Azure also offers finer resource-level breakdowns~\cite{azure-carbon-overview}.
AWS and GCP's monthly allocation rules erase \emph{when} a tenant runs: shifting the same work to cleaner hours does not lower its usage-based share of monthly emissions, even though GCP computes on hourly carbon intensity internally~\cite{aws-ccft-model3,schneider-mattia-2024}.

Providers also differ on \emph{how} Scope~2 is valued: AWS, GCP, and most others expose a location-based figure that reflects the regional grid, so a cleaner region lowers emissions for the same electricity use.
Azure, T~Cloud, and Alibaba Cloud instead publish market-based or ``green-power-adjusted'' figures without a separate location-based Scope~2 view~\cite{azure-carbon-data,tcloud-sustainability-dashboard,alibaba-cloud-carbon-footprint}, obscuring the grid signal behind the region action.

\section{Reconciling Bottom-Up and Top-Down}
\label{sec:rsci}

We introduce \emph{reconciled} \SCI (\RSCI), a family of per-workload metrics that anchors bottom-up carbon estimates to a provider's top-down report.
Its purpose is \emph{attributional}: assigning the reported emissions, including shared overheads, to workloads.
The residual closes the gap between these views, so attributed workload emissions sum to the reported total by construction.
\SCI alone does not guarantee this consistency.
This makes disagreements explicit without treating either estimate as ground truth.

\emph{Consequential} analysis asks how emissions change as a result of an intervention.
El-Zahr and Zilberman~\cite{elzahr2025measurement} make this distinction in accounting for network traffic.
Building on demand-aware attribution in Fair-CO$_2$~\cite{han2025fairco2}, we propose allocating energy- and capacity-driven residual components according to the activities and provisioning requirements that drive them.

\parag{Setup}
Let $i$ index a \emph{workload}: a distinct billable unit a tenant runs, like a service or a batch job.
A reporting \emph{slice} is the provider's finest published granularity for one tenant (service $\times$ region $\times$ month).
A workload can span multiple slices (e.g., multiple services or regions).
Rewriting \OSCI and \SCI using this per-workload notation we get
\begin{equation}
\OSCI_i \;=\; \frac{\sum_s \E_{i,s}\cdot\I_s}{\R_i}
\quad\text{and}\quad
\SCI_i \;=\; \frac{\sum_s \E_{i,s}\cdot\I_s + \M_i}{\R_i},
\label{eq:osci-sci-repetition}
\end{equation}
where $\E_{i,s}$ is the workload's modeled energy within the slice, $\I_s$ the slice's location-based carbon intensity, $\M_i$ the workload machine's embodied term, and $\R_i$ its total functional output over the period.

\OSCI, \SCI, and \RSCI all rely on a bottom-up energy estimate $\E_{i,s}$.
We can, for example, derive $\E_{i,s}$ from per-resource quantities,
\begin{equation}
\E_{i,s} \;=\; \mathrm{PUE}\cdot\sum_{p} q_{i,s,p} \cdot \varepsilon_p,
\label{eq:bu-energy}
\end{equation}
where $p$ indexes resources (vCPU-hours, GPU-hours, GB-hours of storage, I/O ops, network GB),
$q_{i,s,p}$ is workload $i$'s measured consumption of resource $p$ within slice $s$,
$\varepsilon_p$ is a per-type energy coefficient from sources such as BoaviztAPI~\cite{simon2024boaviztapi} or the Cloud Carbon Footprint project~\cite{cloudcarbonfootprint}.
$\mathrm{PUE}$ (power usage effectiveness) is an optional factor for modeling facility overhead.

\parag{Reconciliation}
A tenant's bottom-up model cannot capture all of Scope~2: energy coefficients are approximate, shared idle capacity is unobservable, and a fixed PUE ignores that real facility overhead varies with season and weather.
Emissions in Scope 1 and 3 are even less observable.
Therefore, for each slice, we define the \emph{residual} emissions as the difference between the provider's top-down reported total carbon footprint $C^{\downarrow}_{s}$ and the bottom-up estimate:
\begin{equation}
\Delta_{s} \;=\; C^{\downarrow}_{s} \;-\; \sum_i \E_{i,s}\,\I_s .
\label{eq:residual-total}
\end{equation}
\RSCI distributes this residual across the slice's workloads by weights.
It uses the same operational term as \OSCI, but adds a residual term, taken over one or more \emph{residual components} $\Delta^{c}_s$ (with $\sum_c \Delta^{c}_s = \Delta_s$), each distributed across the slice's workloads by weights $w^{c}_{i,s}$ that sum to one ($\sum_{i \in s} w^{c}_{i,s} = 1$):
\begin{equation}
\RSCI_i \;=\; \frac{\sum_s \bigl( \overbrace{\E_{i,s}\cdot\I_s}^\text{operational term} \;+\; \overbrace{\textstyle\sum_c w^{c}_{i,s}\,\Delta^{c}_{s}}^\text{residual term}\;\;\;\bigr)}{\R_i}.
\label{eq:rsci}
\end{equation}
Attributed carbon then reconciles by construction: scaling each score back to carbon by its $\R_i$, the workload attributions sum to the reported total across slices,
\begin{equation}
\textstyle\sum_i \RSCI_i\,\R_i
=
\textstyle\sum_s
\left(
\sum_{i \in s} \E_{i,s}\I_s
+
\Delta_s
\right)
=
\textstyle\sum_s C^\downarrow_s .
\label{eq:reconcile}
\end{equation}
A workload spanning several slices sums its per-slice attributions, and since each balances on its own, reconciliation holds across slices.
This identity holds for \emph{any} partition of the residual into components and \emph{any} normalized weights.

\parag{Decomposing the residual}
The simplest allocation choice would be to split the full residual $\Delta_s$ evenly, giving each of the slice's $N_s$ workloads a share $w_i = 1/N_s$.
This reconciles, but ignores differences in workloads' energy use and capacity requirements.
To improve this, we first decompose the residual the way the provider reports it, by GHG scope.
Let $C_{s}^{\downarrow\text{S1}}$, $C_{s}^{\downarrow\text{S2}}$, $C_{s}^{\downarrow\text{S3}}$ denote the top-down reported per-slice, per-scope emissions.
Since our bottom-up model estimates only operational Scope~2 carbon, the per-scope residuals (summing to $\Delta_s$) are
\begin{equation}
\Delta^{\text{S1}}_{s} = C^{\downarrow\text{S1}}_{s}, \qquad
\Delta^{\text{S2}}_{s} = C^{\downarrow\text{S2}}_{s} - \sum_i \E_{i,s}\,\I_s, \qquad
\Delta^{\text{S3}}_{s} = C^{\downarrow\text{S3}}_{s},
\label{eq:residual-scopes}
\end{equation}
$\Delta^{\text{S1}}$ absorbs diesel and refrigerants.
$\Delta^{\text{S2}}$ includes unmodeled idle energy, facility overhead, and energy-model differences.
$\Delta^{\text{S3}}$ includes embodied hardware and the rest of Scope~3.

The scope split mirrors the provider's reporting but not the actions a tenant controls.
What matters for incentives is whether an emission scales with \emph{energy usage} (i.e., the bottom-up model) or with \emph{provisioned capacity} (i.e., idle power and embodied emissions).
We therefore propose regrouping the three scope residuals into an energy-driven $\Delta^{\text{en}}_{s}$ and a capacity-driven $\Delta^{\text{cap}}_{s}$ term, writing $\beta^{k}\in[0,1]$ for the energy-driven fraction of scope $k$'s residual:
\begin{equation}
\Delta^{\text{en}}_{s} = \sum_{k} \beta^{k}\,\Delta^{k}_{s}, \qquad
\Delta^{\text{cap}}_{s} = \sum_{k} (1-\beta^{k})\,\Delta^{k}_{s}.
\label{eq:residual-levers}
\end{equation}
The betas must be estimated by tenants from the provider's methodology.
For a positive residual, a larger $\beta$ shifts weight from capacity requirements to energy use.
As a starting point, we treat Scope~1 as capacity-driven ($\beta^{\text{S1}}=0$).
Scope~2 mixes energy-driven terms (unmodeled overhead, energy-model differences) with capacity-driven ones (idle power), so $\beta^{\text{S2}}$ depends on how well the bottom-up model captures the energy-driven part.
Scope~3 combines capacity-driven terms (embodied hardware and buildings) with energy-driven terms (upstream electricity, i.e.\ FERA).

Since $\Delta^{\text{en}}_s + \Delta^{\text{cap}}_s = \Delta_s$, any $\beta$ preserves reconciliation.
The resulting incentives also depend on residual signs, allocation weights, and how the provider's reported total responds to an action.

\parag{Energy allocation}
The energy-driven residual $\Delta^{\text{en}}$ should scale proportionally with the operational term.
At fixed slice intensity $\I_s$, we use the normalized weight
\begin{equation}
w^{\text{en}}_{i,s} = \frac{\E_{i,s}\cdot\I_s}{\sum_{j \in s} \E_{j,s}\cdot\I_s} = \frac{\E_{i,s}}{\sum_{j \in s} \E_{j,s}}.
\label{eq:w-energy}
\end{equation}
With time-varying carbon intensity, replace $\E_{i,s}\I_s$ by time-summed operational emissions in both the bottom-up term and these weights, retaining the reporting slice for reconciliation.

\parag{Capacity allocation}
We propose allocating the capacity-driven residual $\Delta^{\text{cap}}$ according to workloads' provisioning requirements.
For example, (i) demand peaks require providers to hold capacity that is idle outside peak periods, (ii) availability requirements call for redundant capacity~\cite{eilam2021transparent}, and (iii) latency constraints may require warm instances or spare capacity to avoid startup and queuing delays.
Allocation weights should reflect each workload’s contribution to these capacity needs.

Related work provides starting points for defining these weights.
Islam and Ren allocate static non-IT energy equally among active tenants and dynamic energy in proportion to their IT energy use~\cite{islam2016energyaccounting}.
Similarly, Sharma and Fuerst allocate idle energy equally among active functions but control-plane energy by invocation count~\cite{sharma2024accountable_footprint_serverless,sharma2025disaggregation}.
For capacity driven by peaks, Han et al. use Shapley-based attribution to link embodied carbon to peak demand~\cite{han2024gametheoretic,han2025fairco2}, and Schien et al. assign more fixed network energy to traffic during peak periods~\cite{schien2025causal}.
Choosing among such rules requires considering fairness among tenants~\cite{westerhof2023allocation,montoyafranco2025fairwatt}, optimization incentives~\cite{li2025llmserving}, and the stability of demand responses highlighted by peak-based pricing~\cite{nasiriani2016peakpricing}.
Adapting these approaches to residual components and evaluating the resulting incentives remain future work.

\parag{Online use of \RSCI}
Exact \RSCI scores are retrospective as the residual $\Delta_s$ is only known once the provider publishes its top-down report.
During the period, tenants can still optimize against provisional scores based on historical reports and load forecasts.
These scores are then adjusted to match the final \RSCI values once the report arrives.
Validating these estimates and quantifying their uncertainty remain future work.

\section{Example}
\label{sec:demo}

\begin{figure}[t]
\centering
\includegraphics[width=\columnwidth]{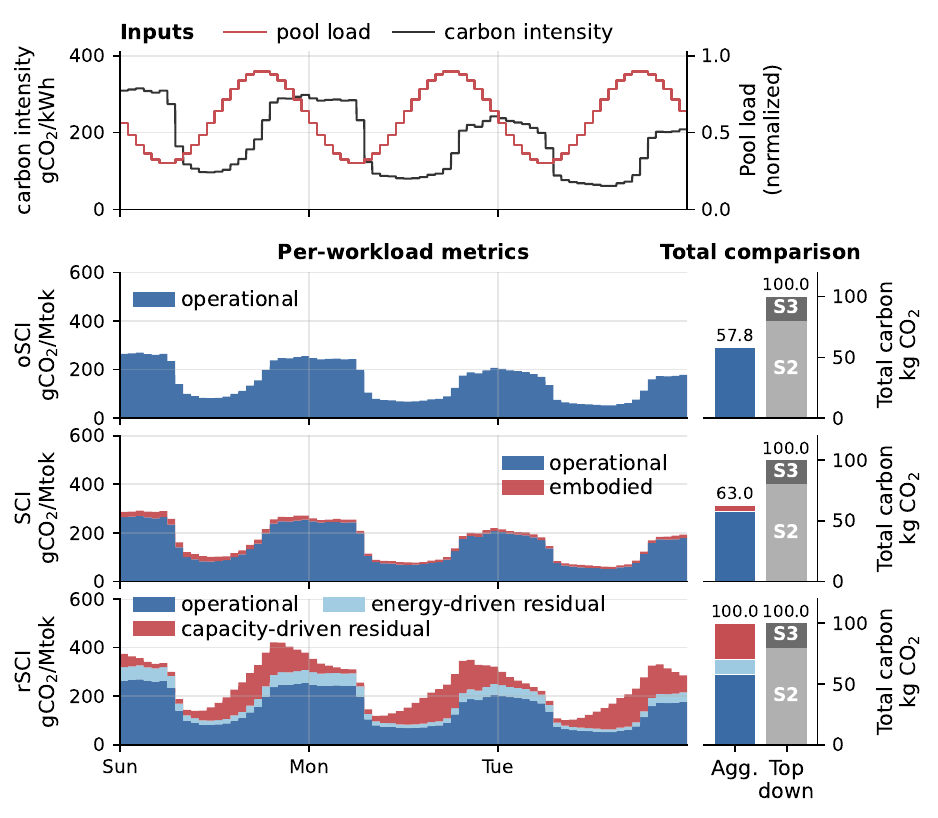}
\vspace{-7mm}
\caption{Comparison of different carbon-per-token metrics (\OSCI, \SCI, \RSCI) over time, reported hourly.
The bar chart compares each metric's aggregated total against the provider's top-down report.}
\vspace{-3mm}
\label{fig:demo}
\end{figure}

We illustrate \RSCI's reconciliation and incentives.
Using an assumed provider report, we model a single workload on a single node, reporting carbon per token served. Results are shown in Figure~\ref{fig:demo}.
Code and data for reproducing the experiment and Figure~\ref{fig:demo} are available at \url{https://github.com/birnbaum/rsci-example}.

\parag{Scenario}
We model an internal LLM service on an 8$\times$H100 node over three days, reporting hourly emissions in gCO$_2$ per million tokens (gCO$_2$/Mtok).
We model power as linear between $2.5$\,kW idle and the rated maximum of $10.2$\,kW~\cite{nvidia-dgx-h100-datasheet} and we assume a PUE of $1.2$.
We model load using the 2024 Azure LLM inference trace~\cite{stojkovic2025dynamollm} and carbon intensity using California (CAISO) data from Electricity Maps~\cite{electricitymaps2026}.
The shared pool follows a synthetic daily load cycle of approximately 30--90\,\% of available capacity, independent of the workload trace.
Retrospectively, the provider reports the node's footprint top-down: $C^{\downarrow\text{S2}}\!=\!80$\,kg CO$_2$ for Scope~2 and $C^{\downarrow\text{S3}}\!=\!20$\,kg for Scope~3 (we omit Scope~1 for simplicity).
We ask how well the aggregated carbon-per-token metrics reconcile to that $100$\,kg total, and what behaviour they reward along the way.

\parag{\OSCI}
This operational metric counts only the marginal energy the work draws above idle, scaled by the PUE and the hourly carbon intensity.
It rewards the main actions a tenant controls in this scenario: using less energy (\E) and computing work during cleaner hours (\I).
By construction the node's idle power and its embodied hardware are not considered.
Lacking any embodied term, it cannot be misguided by depreciated hardware (i.e., the sunk-carbon fallacy~\cite{bashir2024}).
\OSCI reports between 52 and 270\,gCO$_2$/Mtok (${\sim}5.2\times$ range) and attributes about 58\,kg in total.

\parag{\SCI}
\SCI adds an amortized embodied term to the same operational energy estimate used for \OSCI.\footnote{We assume the node's cradle-to-gate embodied carbon is $2500$\,kg: NVIDIA reports $1312$\,kg for the 8-GPU HGX~H100 baseboard~\cite{nvidia2025hgxh100pcf} alone. We assume the node is amortised over a 4-year lifespan, giving $5.1$\,kg over three days, equivalent to $12.4$\,gCO$_2$/Mtok across the $415.4$ million tokens served.}
Over the three days, this adds $5.1$\,kg, bringing \SCI to $63.0$\,kg, which still falls short of the provider's $100$\,kg report.

\parag{\RSCI}
\RSCI keeps \OSCI's operational term and adds the residual, the gap between the provider's reported total and the bottom-up estimate.
Following \S\ref{sec:rsci}, we split the residual into an energy-driven part, $\Delta_s^{\text{en}}\!\approx\!12.2$\,kg, and a capacity-driven part, $\Delta_s^{\text{cap}}\!\approx\!29.9$\,kg.
For illustration, we use $\beta^{\text{S2}}\!=\!0.1$ and $\beta^{\text{S3}}\!=\!0.5$.

The energy-driven residual is allocated in proportion to modeled operational emissions.
For the capacity-driven residual, we use a simple rule inspired by Schien et al.~\cite{schien2025causal} that gives more weight to work during busy pool hours.
The hourly weights are $Q_h s_h^2/\sum_t Q_t s_t^2$, where $Q_h$ is tokens served and $s_h$ is aggregate pool demand divided by fixed available capacity.

The energy-driven residual adds 11--57\,gCO$_2$/Mtok (${\sim}5.2\times$ range).
The capacity-driven residual contributes $16$--$139$\,gCO$_2$/Mtok, compared with $72$ under uniform allocation per token.
By construction, multiplying each hourly \RSCI score by its token count and summing gives the reported $100$\,kg.
Compared with uniform allocation, this rule attributes more capacity emissions per token during shared peaks.
Shifting flexible work to quieter hours can reduce future capacity needs when peaks determine provisioning.

\begin{table}[t]
\centering
\caption{Properties of per-workload carbon metrics. \faithful~supported, \conditional~conditional (discussed below), \notrefl~not supported.}
\vspace{-3mm}
\label{tab:properties}
\small
\setlength{\tabcolsep}{6pt}
\renewcommand{\arraystretch}{1.2}
\begin{tabular}{@{}l ccc@{}}
\toprule
\textbf{Property} & \textbf{\SCI}~\cite{sci-spec} & \textbf{\OSCI}~\cite{bashir2024} & \textbf{\RSCI} \\
\midrule
Rewards operational actions & \faithful & \faithful & \conditional \\
Attributes embodied carbon & \faithful & \notrefl & \faithful \\
Avoids the sunk-carbon fallacy~\cite{bashir2024} & \notrefl & \faithful & \conditional \\
Allocates residuals by their drivers & \notrefl & \notrefl & \conditional \\
Reconciles to cloud provider's report & \notrefl & \notrefl & \faithful \\
\bottomrule
\end{tabular}
\vspace{-2mm}
\end{table}

\parag{Comparison}
Table~\ref{tab:properties} summarizes the accounting properties and conditional incentives of \RSCI:
\begin{itemize}
\setlength{\itemsep}{2pt}
  \item \emph{Rewards operational actions.} Its operational term responds to energy savings and cleaner hours or regions.
  These improvements reduce the reconciled total \emph{if the provider's report reflects them}.
  Appendix~\ref{app:operational-incentives} gives the conditions for individual workload incentives.
  \item \emph{Attributes embodied carbon.} Past hardware purchases are a sunk cost and should not steer operational choices~\cite{bashir2024}, but ignoring embodied carbon omits part of the reported footprint and drops an incentive to provision less hardware.
  \item \emph{Avoids the sunk-carbon fallacy~\cite{bashir2024}.} Allocation by energy and capacity requirements gives no discount for depreciated hardware within a reporting slice.
  Fleet age can still affect the provider's inventory.
  \item \emph{Allocates residuals by their drivers.} We propose allocating energy residuals by operational emissions and capacity residuals by provisioning needs.
  Defining and validating suitable capacity weights remains future work.
  The incentive conditions are given in Appendix~\ref{app:operational-incentives}.
  \item \emph{Reconciles by construction.} Of the three metrics, only \RSCI guarantees that attributed workload emissions sum to the provider's reported total.
\end{itemize}

\section{Barriers to Practical Adoption}
\label{sec:practice}

Reconciliation alone is not sufficient for trustworthy accounting.
In practice, computing credible \RSCI scores requires provider reports that are sufficiently detailed, transparent, and standardized.

To evaluate whether this is possible today, we attempted to compute \RSCI scores using production accounts on AWS and Azure, as well as test deployments on Scaleway.
We identify three barriers to obtaining the information needed for reconciliation.

\parag{Top-down reports are too coarse}
\RSCI's per-workload attribution is straightforward when a reporting slice contains a single workload, as in \S\ref{sec:demo}.
In practice, however, providers aggregate many workloads into a small number of monthly service categories.
The coarser the report, across both resources and time, the more attribution depends on heuristic allocation that cannot be checked against the reported data.

\begin{figure}[t]
\centering

  \IfFileExists{figures/aws_spend_carbon.pdf}{
  \includegraphics[width=\columnwidth]{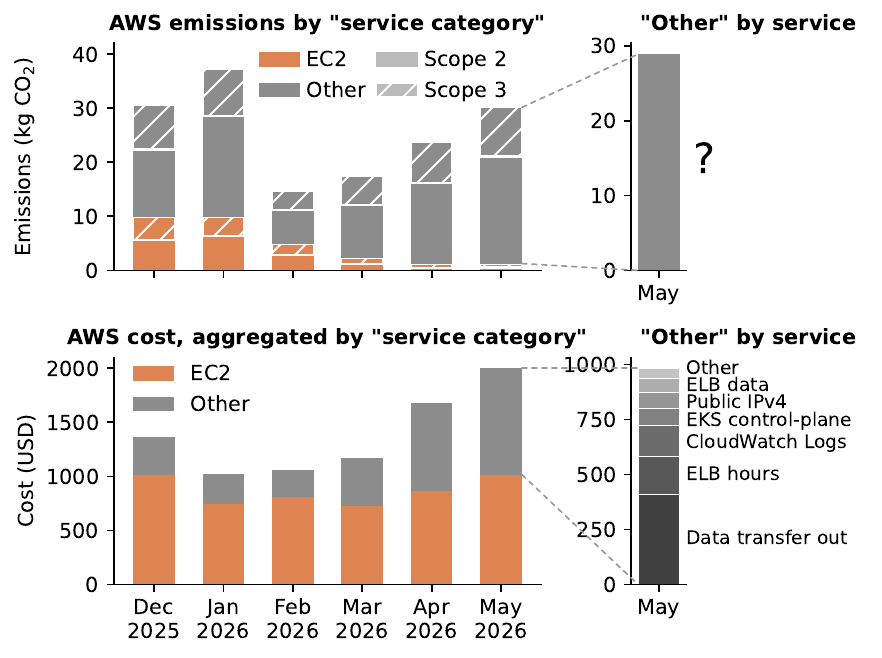}
}{\fbox{\parbox[c][3cm][c]{.9\columnwidth}{\centering\small aws\_spend\_carbon.pdf\\(pending notebook rework)}}}
 \vspace{-7mm}
\caption{Cloud emissions and cost reported for an AWS production account.
While billing is granular and can be traced to usage, the carbon report is coarse and opaque.}
\label{fig:aws}
\vspace{-2mm}
\end{figure}

We observed this on six months of production data from an AWS account running a web service in Frankfurt (December~2025--May~2026).
Although AWS bills individual services separately, its carbon report aggregates most emissions into a service category called ``Other'', see Figure~\ref{fig:aws}.
Across the six months, this category accounts for 67--96\,\% of reported emissions and combines diverse services such as load balancing, data transfer, control-plane operations, and public IPv4 addresses.
The report provides no resource-level breakdown to validate attribution to individual workloads.

\begin{callforaction}
Providers already meter and bill customer usage at fine granularity.
Customer-facing carbon reporting should preserve this and publish emissions per billable resource and reporting period.
Reports should also distinguish energy and capacity components.
Allocation by shared peaks requires demand data for the shared resource pool.
\end{callforaction}

\parag{Reported totals are opaque}
Even at sufficient granularity, a reconciled metric can only anchor to a provider's reported total if that total is itself trustworthy.
Today, customer-facing reports are difficult to reproduce or interpret, making it impossible to distinguish modeling error from reporting methodology~\cite{eilam2021transparent}.

We found this even for AWS EC2, whose methodology is comparatively well documented.
Using standard bottom-up energy models, our estimated emissions were roughly 2$\times$ the reported location-based Scope~2 total at idle and 5$\times$ at moderate utilization, producing a negative residual.
A negative residual is mathematically admissible, but without further information we cannot distinguish model error from differences in reporting boundaries or allocation.
Allocating a negative capacity residual by peak contribution gives larger contributors a larger credit, reversing the intended incentive.

Other providers exhibit similar issues.
In our June~2026 Scaleway tests, daily emissions totals changed when we queried them a week later, and included emissions from before account creation.
Microsoft provides stable monthly reports, but explicitly notes that customer-facing emissions ``will differ from those reflected in Microsoft's corporate disclosure''~\cite{microsoft-s3-2021}.
Without stable, reproducible totals, reconciliation cannot provide trustworthy workload attribution.
Man and Kleppmann~\cite{man2026emissionimpossible} demonstrate how customer emissions reports can be verified automatically at cloud scale without exposing confidential inputs.

\begin{callforaction}
Customer-facing carbon reports should be reproducible and auditable.
Providers should disclose enough intermediate information and methodology for tenants to independently explain reported totals and distinguish reporting assumptions from modeling error.
\end{callforaction}

\parag{There is no common standard}
As we outlined in \S\ref{sec:sota}, providers differ in what they report, how they allocate it, and what they count.
Azure and T~Cloud expose only market-based Scope~2 emissions, leaving no location-based figure to reconcile against.
AWS allocates by usage or equivalent revenue, GCP incorporates list-price weighting, while Oracle, Scaleway, and T~Cloud use hardware energy measurements or power models whose customer allocation rules differ.
Even the hyperscalers disagree on Scope~3 boundaries, including different combinations of upstream fuel losses, buildings, end-of-life, and business travel.

These differences ultimately reflect gaps in today's GHG Protocol guidance for cloud infrastructure, which leaves shared-infrastructure allocation, equipment lifetimes, and several reporting boundaries largely to provider discretion.
\RSCI depends only on the structure of top-down reporting rather than any specific methodology, so improvements to the standard only strengthen the metric.

\begin{callforaction}
The GHG Protocol should define cloud-specific guidance for customer-facing reporting, including common reporting boundaries, allocation rules for shared infrastructure, and standardized outputs.
A standard would make provider reports comparable and enable \RSCI to be portable across clouds.
\end{callforaction}

\section{Conclusion}
\label{sec:future}

In this paper, we outlined a vision for reconciling bottom-up workload metrics with providers' top-down carbon reports.
Our proposed \RSCI framework allocates the residual so that workload emissions sum to the reported total.
Decomposing the residual into energy- and capacity-driven components can preserve operational incentives while attributing idle capacity and embodied carbon to the workloads that drive them.
However, we find that today's provider reports remain too coarse, opaque, and inconsistent to support credible reconciliation in practice.

Future work should evaluate \RSCI on production workloads across clouds, validate allocation weights and their incentives under different residual signs, betas, and provider responses, and aim to build reproducible reconciliation pipelines.
Moreover, we should explore online proxies that estimate residuals and capacity weights from historical reports and load forecasts, and expose the resulting uncertainty, compounded by the inherent uncertainty of embodied-carbon models~\cite{bhagavathula2024uncertainty}, until the final top-down report arrives.
With more granular and transparent reporting as well as clearer cloud reporting standards, \RSCI could make carbon accountability as operational as cost accountability is today.

\begin{acks}
We thank the anonymous SoCC reviewers and our shepherd for their constructive feedback, which helped clarify the scope and incentive properties of reconciled SCI.
L.L. was supported by the Wellcome Trust [313909/Z/24/Z] and by core funding from the British Heart Foundation (RG/F/23/110103), NIHR Cambridge Biomedical Research Centre (NIHR203312)\footnote{The views expressed are those of the authors and not necessarily those of the NIHR or the Department of Health and Social Care.}, BHF Chair Award (CH/12/2/29428), and by Health Data Research UK (HDRUK2023.0028), which is funded by the UK Medical Research Council, Engineering and Physical Sciences Research Council, Economic and Social Research Council, Department of Health and Social Care (England), Chief Scientist Office of the Scottish Government Health and Social Care Directorates, Health and Social Care Research and Development Division (Welsh Government), Public Health Agency (Northern Ireland), British Heart Foundation and the Wellcome Trust.
\end{acks}

\bibliographystyle{ACM-Reference-Format}
\bibliography{references}

\clearpage
\appendix
\onecolumn
\input{appendix-survey}
\clearpage
\input{appendix-incentives}

\end{document}

%% file: appendix-survey.tex
\begingroup
\definecolor{surveyHeading}{HTML}{285A8E}
\definecolor{surveyReport}{HTML}{12633D}
\definecolor{surveyMetrics}{HTML}{8A6800}
\definecolor{surveyClaims}{HTML}{A85516}
\definecolor{surveyMissing}{HTML}{9B4141}
\newcommand{\surveyclass}[2]{\textcolor{#1}{\textbf{#2}}}
\newcommand{\surveysource}[2]{#2~\cite{#1}}
\color{black}
\hypersetup{urlcolor=black,linkcolor=black,citecolor=black}
\newcommand{\surveygroup}[1]{\colorbox{surveyHeading!10}{\parbox{\dimexpr\textwidth-2\fboxsep\relax}{\textcolor{surveyHeading}{\strut\textbf{#1}}}}}
\section{Provider Survey}
\label{app:provider-survey}
This appendix documents the provider survey discussed in Section~\ref{sec:sota}.

\subsection{Survey Scope and Provider Disclosures}
\label{app:inventory}
We reviewed publicly available environmental disclosures from 50 cloud providers across general-purpose compute, GPU/HPC (neocloud), CDN/edge, application hosting, and object storage.
The sample includes major cloud platforms and selected regional and specialized providers, with a focus on Europe and North America.
We excluded colocation-only providers, general business SaaS, model-only APIs, and independent carbon-accounting tools.
Providers were selected regardless of their environmental disclosures.
We initially reviewed official documentation and reports on 27 September 2026.

\smallskip
\noindent\textbf{Classification.} Categories reflect providers' publicly available documentation, reports, and claims.
We did not test every customer reporting tool or independently verify all reported figures.
Each provider is assigned the first applicable category below.
\begin{itemize}
\setlength{\itemsep}{1pt}
\setlength{\parskip}{0pt}
\item \textcolor{surveyReport}{\textbf{Customer report}} (15): emissions attributed to an individual customer's usage, excluding reports that show only estimated savings.
\item \textcolor{surveyMetrics}{\textbf{Public metrics}} (17): quantified corporate or operational environmental results, without a per-customer report.
\item \textcolor{surveyClaims}{\textbf{Claims only}} (12): environmental commitments, practices, targets, or projected savings, without quantified results.
\item \textcolor{surveyMissing}{\textbf{Not found}} (6): no qualifying disclosure identified in the reviewed sources.
\end{itemize}

\small
\renewcommand{\arraystretch}{1.20}
\setlength{\tabcolsep}{4pt}
\begin{longtable}{@{}p{.22\textwidth}p{.24\textwidth}p{\dimexpr.54\textwidth-16pt\relax}@{}}
\caption{Environmental disclosures from the surveyed cloud providers.}
\vspace{-3mm}
\label{tab:provider-inventory}\\
\toprule
\textbf{Provider} & \textbf{Disclosure} & \textbf{Source} \\
\midrule
\endfirsthead
\multicolumn{3}{l}{\emph{Table~\ref{tab:provider-inventory}: provider disclosures (continued)}} \\
\toprule
\textbf{Provider} & \textbf{Disclosure} & \textbf{Source} \\
\midrule
\endhead
\midrule
\multicolumn{3}{r}{\emph{Continued on next page}} \\
\endfoot
\bottomrule
\endlastfoot
\multicolumn{3}{@{}l@{}}{\surveygroup{Hyperscalers (9)}} \\*
Alibaba Cloud & \surveyclass{surveyReport}{Customer report} & \surveysource{alibaba-cloud-carbon-footprint}{Cloud product carbon footprint documentation} \\*
AWS & \surveyclass{surveyReport}{Customer report} & \surveysource{aws-sustainability-concepts}{Sustainability console documentation} \\*
Azure & \surveyclass{surveyReport}{Customer report} & \surveysource{azure-carbon-overview}{Azure Carbon Optimization overview} \\*
GCP & \surveyclass{surveyReport}{Customer report} & \surveysource{google-carbon-methodology}{Carbon Footprint methodology} \\*
IBM Cloud & \surveyclass{surveyReport}{Customer report} & \surveysource{ibm-carbon-current}{Working with IBM Cloud's carbon calculator} \\*
Oracle (OCI) & \surveyclass{surveyReport}{Customer report} & \surveysource{oracle-emissions-management}{OCI Emissions Management documentation} \\*
Baidu AI Cloud & \surveyclass{surveyMetrics}{Public metrics} & \surveysource{baidu-esg-2024}{Baidu ESG report (2024)} \\*
Huawei Cloud & \surveyclass{surveyMetrics}{Public metrics} & \surveysource{huawei-green-data-centers}{Huawei Cloud: green data centers and PUE} \\*
Tencent Cloud & \surveyclass{surveyMetrics}{Public metrics} & \surveysource{tencent-carbon-neutrality}{Tencent climate disclosures} \\
\multicolumn{3}{@{}l@{}}{\surveygroup{Other general-purpose cloud providers (18)}} \\*
Exoscale & \surveyclass{surveyReport}{Customer report} & \surveysource{exoscale-environmental-impact}{Environmental Impact reporting documentation} \\*
Infomaniak & \surveyclass{surveyReport}{Customer report} & \surveysource{infomaniak-invoice-footprint}{Invoice-footprint methodology (FAQ 1160)} \\*
OUTSCALE & \surveyclass{surveyReport}{Customer report} & \surveysource{outscale-carbon-reporting}{Carbon Footprint reporting documentation} \\*
OVHcloud & \surveyclass{surveyReport}{Customer report} & \surveysource{ovh-eit-v2}{Environmental Impact Tracker methodology} \\*
Scaleway & \surveyclass{surveyReport}{Customer report} & \surveysource{scaleway-footprint-api}{Environmental Footprint API documentation} \\*
T Cloud Public & \surveyclass{surveyReport}{Customer report} & \surveysource{tcloud-sustainability-dashboard}{Sustainability Dashboard and methodology} \\*
GleSYS & \surveyclass{surveyMetrics}{Public metrics} & \surveysource{glesys-sustainability-2024}{Sustainability report announcement (2024)} \\*
Hetzner & \surveyclass{surveyMetrics}{Public metrics} & \surveysource{hetzner-environmental-statement-2025}{Environmental statement (2025, German operations)} \\*
IONOS & \surveyclass{surveyMetrics}{Public metrics} & \surveysource{ionos-environment}{Environmental disclosures} \\*
JD Cloud & \surveyclass{surveyMetrics}{Public metrics} & \surveysource{jd-esg-2024}{JD.com ESG report announcement (2024)} \\*
Rackspace & \surveyclass{surveyMetrics}{Public metrics} & \surveysource{rackspace-sustainability-2025}{Sustainability report announcement (2025)} \\*
STACKIT & \surveyclass{surveyMetrics}{Public metrics} & \surveysource{stackit-colocation}{Colocation specifications and PUE} \\*
UpCloud & \surveyclass{surveyMetrics}{Public metrics} & \surveysource{upcloud-esg-2024}{ESG report (2024)} \\*
Yandex Cloud & \surveyclass{surveyMetrics}{Public metrics} & \surveysource{yandex-sustainability-2024}{Yandex sustainability report (2024)} \\*
Civo & \surveyclass{surveyClaims}{Claims only} & \surveysource{civo-uk-ai-sovereignty}{UK infrastructure and efficiency commitments} \\*
Leafcloud & \surveyclass{surveyClaims}{Claims only} & \surveysource{leafcloud-gpu-carbon-calculator}{GPU carbon savings calculator explained} \\*
Vultr & \surveyclass{surveyClaims}{Claims only} & \surveysource{vultr-sabey-expansion}{Sabey partnership and environmental commitments} \\*
DigitalOcean & \surveyclass{surveyMissing}{Not found} & -- \\
\multicolumn{3}{@{}l@{}}{\surveygroup{GPU/HPC providers (10)}} \\*
atNorth & \surveyclass{surveyMetrics}{Public metrics} & \surveysource{atnorth-sustainability-2024}{Sustainability report announcement (2024)} \\*
Crusoe & \surveyclass{surveyMetrics}{Public metrics} & \surveysource{crusoe-impact-2024}{Impact report (2024)} \\*
Nebius & \surveyclass{surveyMetrics}{Public metrics} & \surveysource{nebius-sustainability}{Sustainability disclosures} \\*
Verda & \surveyclass{surveyMetrics}{Public metrics} & \surveysource{verda-locations-sustainability}{Locations and sustainability documentation} \\*
CoreWeave & \surveyclass{surveyClaims}{Claims only} & \surveysource{coreweave-10k-2025}{Form 10-K, fiscal year 2025} \\*
Hyperstack & \surveyclass{surveyClaims}{Claims only} & \surveysource{hyperstack-infrastructure}{Infrastructure and renewable-power claims} \\*
Lambda & \surveyclass{surveyClaims}{Claims only} & \surveysource{lambda-hydrogen-gb300}{Hydrogen-powered facility announcement} \\*
Nscale & \surveyclass{surveyClaims}{Claims only} & \surveysource{nscale-sustainable-infrastructure}{Sustainable AI infrastructure commitments} \\*
RunPod & \surveyclass{surveyClaims}{Claims only} & \surveysource{runpod-legal-compliance}{Environmental procurement criteria} \\*
Paperspace & \surveyclass{surveyMissing}{Not found} & -- \\
\multicolumn{3}{@{}l@{}}{\surveygroup{CDN/edge providers (6)}} \\*
Akamai & \surveyclass{surveyReport}{Customer report} & \surveysource{akamai-customer-engagement}{Customer emissions reporting and calculator} \\*
Fastly & \surveyclass{surveyReport}{Customer report} & \surveysource{fastly-sustainability-output}{Sustainability dashboard documentation} \\*
CDNetworks & \surveyclass{surveyMetrics}{Public metrics} & \surveysource{cdnetworks-sustainability}{Sustainability metrics} \\*
Cloudflare & \surveyclass{surveyMetrics}{Public metrics} & \surveysource{cloudflare-impact-2025}{Impact report (2025): corporate emissions inventory} \\*
Gcore & \surveyclass{surveyClaims}{Claims only} & \surveysource{gcore-infrastructure}{Infrastructure and efficiency claims} \\*
Bunny.net & \surveyclass{surveyMissing}{Not found} & -- \\
\multicolumn{3}{@{}l@{}}{\surveygroup{Application hosting platforms (5)}} \\*
Netlify & \surveyclass{surveyClaims}{Claims only} & \surveysource{netlify-sustainability}{Sustainability commitments} \\*
Vercel & \surveyclass{surveyClaims}{Claims only} & \surveysource{vercel-avanade-case-study}{Customer case study: website carbon reduction} \\*
Fly.io & \surveyclass{surveyMissing}{Not found} & -- \\*
Railway & \surveyclass{surveyMissing}{Not found} & -- \\*
Render & \surveyclass{surveyMissing}{Not found} & -- \\
\multicolumn{3}{@{}l@{}}{\surveygroup{Object storage providers (2)}} \\*
Wasabi & \surveyclass{surveyReport}{Customer report} & \surveysource{wasabi-carbon-subaccounts}{Account Control Manager: Carbon Footprint} \\*
Backblaze & \surveyclass{surveyClaims}{Claims only} & \surveysource{backblaze-infrastructure}{Infrastructure, recycling, and cooling practices} \\
\end{longtable}
\normalsize

\clearpage
\subsection{Coverage and Granularity of Cloud Emissions Reports}
\label{app:reporting-coverage}
Of the 50 surveyed providers, 15 claim to offer reports that attribute emissions to individual customers.
Table~\ref{tab:reporting-coverage} compares their reporting intervals, the breakdowns available to customers, and the emissions sources and electricity accounting methods covered by their reports.
The entries are based on the public documentation cited beside each provider. We additionally checked the documentation for reporting breakdowns on 6 October 2026.
The columns should be read as follows:

\begin{itemize}
\item \textbf{Reporting interval} gives the shortest period for which a provider documents emissions totals. We did not test all dashboards, but daily reporting in particular warrants caution. For example, in our June~2026 tests, Scaleway's daily reports appeared unreliable.
\item \textbf{Reporting breakdown} lists how customers can split or filter reported emissions. Detail varies by service and reporting interface. The listed dimensions are not necessarily available together.
\item The \textbf{Scopes 1, 2, and 3} columns show which emissions providers account for when calculating the footprints allocated to customers. Scope~1 covers direct emissions, such as fuel combustion. The two Scope~2 columns distinguish methods for calculating emissions from electricity use. \emph{Location-based} estimates use the average emissions per unit of electricity supplied by the local grid. \emph{Market-based} estimates use emission factors associated with the provider's electricity purchases, including contracts and renewable energy certificates. These columns also cover electricity classified as provider Scope~3, such as colocation electricity. The five Scope~3 columns compare coverage of selected emissions sources. They do not correspond one-to-one to GHG Protocol categories.
\end{itemize}

We did not independently verify the reported emissions.
Coverage may vary by service and region.
A checkmark indicates inclusion but does not imply complete coverage of an emissions source.
Customer allocation methods and their responses to engineering actions are assessed separately in Table~\ref{tab:levers} and Appendix~\ref{app:action-evidence}.

\begin{table}[h!]
\centering
\caption{Coverage and granularity of cloud emissions reports. \faithful: emissions source included or electricity accounting basis used, \notrefl: source excluded or basis not used, \undocumented: undocumented.}
\vspace{-3mm}
\label{tab:reporting-coverage}
\small
\setlength{\tabcolsep}{2.5pt}
\renewcommand{\arraystretch}{1.15}
\begin{tabularx}{\textwidth}{@{}p{2.45cm}p{1.35cm}Lcccccccc@{}}
\toprule
& & & \textbf{Scope 1} & \multicolumn{2}{c}{\textbf{Scope 2}} & \multicolumn{5}{c}{\textbf{Scope 3}} \\
\cmidrule(lr){4-4}\cmidrule(lr){5-6}\cmidrule(l){7-11}
\textbf{Provider / source} & \makecell[l]{\textbf{Reporting}\\\textbf{interval}} & \textbf{Reporting breakdown} & & \makecell{\textbf{Location-}\\\textbf{based}} & \makecell{\textbf{Market-}\\\textbf{based}} & \makecell{\textbf{Hard-}\\\textbf{ware}} & \makecell{\textbf{Build-}\\\textbf{ings}} & \makecell{\textbf{Upstream}\\\textbf{energy}} & \makecell{\textbf{Trans-}\\\textbf{port}} & \makecell{\textbf{End}\\\textbf{of life}} \\
\midrule
\rowcolor{surveyHeading!10}\multicolumn{11}{l}{\textcolor{surveyHeading}{\strut\textbf{Hyperscalers (6)}}} \\
Alibaba Cloud~\cite{alibaba-cloud-carbon-footprint} & Month & Account & \faithful & \notrefl & \faithful & \notrefl & \notrefl & \notrefl & \notrefl & \notrefl \\
AWS~\cite{aws-ccft-model3,aws-sustainability-concepts} & Month & Account, service category, region & \faithful & \faithful & \faithful & \faithful & \faithful & \faithful & \faithful & \notrefl \\
Azure~\cite{azure-methodology-2026,azure-carbon-overview} & Month & Subscription, resource group, resource type, resource, region & \faithful & \notrefl & \faithful & \faithful & \notrefl & \notrefl & \faithful & \faithful \\
GCP~\cite{google-carbon-methodology,google-carbon-schema} & Month & Billing account, project, service, region or zone & \faithful & \faithful & \faithful & \faithful & \faithful & \faithful & \faithful & \notrefl \\
IBM Cloud~\cite{ibm-carbon-calc-v3,ibm-carbon-current} & Month & Account, resource group, service, region or datacenter & \notrefl & \faithful & \faithful & \notrefl & \notrefl & \notrefl & \notrefl & \notrefl \\
Oracle (OCI)~\cite{oracle-emissions-management,oracle-carbon-viewreports} & Day & Tenant, compartment, tag, service, SKU, resource, region or zone & \faithful & \faithful & \faithful & \notrefl & \notrefl & \notrefl & \notrefl & \notrefl \\
\rowcolor{surveyHeading!10}\multicolumn{11}{l}{\textcolor{surveyHeading}{\strut\textbf{Other general-purpose cloud providers (6)}}} \\
Exoscale~\cite{exoscale-impact-methodology,exoscale-environmental-impact} & Day & Organization, resource type and size, zone & \notrefl & \faithful & \notrefl & \faithful & \notrefl & \faithful & \faithful & \faithful \\
Infomaniak~\cite{infomaniak-invoice-footprint} & Year & Not specified (web hosting/mail) & \notrefl & \notrefl & \faithful & \faithful & \notrefl & \undocumented & \faithful & \faithful \\
OUTSCALE~\cite{outscale-carbon-methodology,outscale-carbon-reporting} & Month & Account, service category, region & \notrefl & \faithful & \notrefl & \faithful & \undocumented & \undocumented & \undocumented & \undocumented \\
OVHcloud~\cite{ovh-eit-v2} & Month & Service, product family, datacenter & \faithful & \faithful & \faithful & \faithful & \faithful & \faithful & \faithful & \undocumented \\
Scaleway~\cite{scaleway-footprint-calculator,scaleway-footprint-api} & Day & Project, service category, product type, SKU, region or zone & \undocumented & \faithful & \notrefl & \faithful & \faithful & \faithful & \faithful & \faithful \\
T Cloud Public~\cite{tcloud-sustainability-dashboard} & Day & Tenant, service & \faithful & \notrefl & \faithful & \notrefl & \notrefl & \notrefl & \notrefl & \notrefl \\
\rowcolor{surveyHeading!10}\multicolumn{11}{l}{\textcolor{surveyHeading}{\strut\textbf{CDN/edge providers (2)}}} \\
Akamai~\cite{akamai-emissions-policy-2025,akamai-customer-reporting} & Quarter & Customer total (finer breakdown unclear) & \faithful & \faithful & \faithful & \faithful & \faithful & \faithful & \faithful & \undocumented \\
Fastly~\cite{fastly-sustainability-output} & Day & Service category (Delivery, Compute, shared components), country & \notrefl & \faithful & \faithful & \notrefl & \notrefl & \faithful & \notrefl & \notrefl \\
\rowcolor{surveyHeading!10}\multicolumn{11}{l}{\textcolor{surveyHeading}{\strut\textbf{Object storage providers (1)}}} \\
Wasabi~\cite{wasabi-carbon-footprint-dashboard,wasabi-carbon-subaccounts} & Month & Account, subaccount & \undocumented & \undocumented & \undocumented & \undocumented & \undocumented & \undocumented & \undocumented & \undocumented \\
\bottomrule
\end{tabularx}

\end{table}

\clearpage
\subsection{Interpreting Table~\ref{tab:levers}}
\label{app:action-evidence}
The following paragraphs explain the five action classifications in Table~\ref{tab:levers}, which covers the 11 providers offering customer reports for general-purpose compute.

\renewcommand{\parag}[1]{\vspace{2mm}\noindent\textbf{#1}\hspace{2mm}}

\parag{Efficient resource: Does using more energy-efficient hardware for the same work lower reported emissions?}
Most methods use prices or pooled resource weights, which need not track the efficiency of the hardware running a workload.
GCP weights SKU usage by list price~\cite{schneider-mattia-2024}, Azure uses cost normalized to exclude discounts~\cite{microsoft-s3-2021,azure-methodology-2026}, and Oracle's spend-based method uses undiscounted charges~\cite{oracle-emissions-management}.
AWS EC2 allocates service totals in proportion to normalized instance-hours~\cite{aws-ccft-model3}, and IBM VPC uses vCPU-hours~\cite{ibm-carbon-calc-v3}, without retaining the energy use of the customer's particular machine.
Alibaba allocates by cloud resource quantity without specifying how its weights account for hardware efficiency~\cite{alibaba-cloud-carbon-footprint}.
OVHcloud models dedicated servers individually but averages hardware impacts within each shared-instance range~\cite{ovh-eit-v2}, while Exoscale averages across each product's hardware pool~\cite{exoscale-impact-methodology}.
Scaleway models CPU-instance power using processor characteristics and CPU utilization, but uses average consumption estimates for GPU instances~\cite{scaleway-instances-footprint}.
T~Cloud maps services to infrastructure components using power profiles derived from measurements, so its estimates depend on the modeled configuration~\cite{tcloud-sustainability-release,tcloud-sustainability-dashboard}.
OUTSCALE does not disclose its customer allocation weights~\cite{outscale-carbon-methodology}.

\parag{Lower utilization: Does lower utilization reduce reported emissions when reserved capacity and runtime stay the same?}
Scaleway's CPU-instance reports use actual CPU utilization in the power model, so reduced activity can lower attributed electricity use, whereas its GPU estimates do not follow customer activity~\cite{scaleway-instances-footprint}.
For AWS EC2~\cite{aws-ccft-model3}, GCP Compute Engine~\cite{schneider-mattia-2024}, IBM VPC~\cite{ibm-carbon-calc-v3} and Exoscale~\cite{exoscale-impact-methodology}, lower CPU activity leaves the reserved quantities used for allocation unchanged.
Azure's normalized-cost weight likewise stays unchanged for the same VM and runtime~\cite{microsoft-s3-2021,azure-methodology-2026}.
OVHcloud estimates server energy at full load and allocates VM footprints by size and billed hours, without using the customer's actual CPU activity~\cite{ovh-eit-v2}.
T~Cloud's release notes describe allocation by provisioned resources~\cite{tcloud-sustainability-release}, while its blog describes allocation by actual utilization~\cite{tcloud-sustainability-dashboard}, without explaining how the two are combined.
Oracle~\cite{oracle-emissions-management}, Alibaba~\cite{alibaba-cloud-carbon-footprint} and OUTSCALE~\cite{outscale-carbon-methodology} do not specify whether customer CPU activity changes the allocation at a fixed reservation.

\parag{Right-size: Does reserving less capacity for the same work and runtime lower reported emissions?}
Smaller reservations reduce the quantities used for allocation in most methods.
Within a comparable instance family, this reduces normalized instance-hours for AWS EC2~\cite{aws-ccft-model3}, billable resource units for GCP~\cite{schneider-mattia-2024}, vCPU-hours for IBM VPC~\cite{ibm-carbon-calc-v3}, and reserved-resource shares for OVHcloud~\cite{ovh-eit-v2} and Exoscale~\cite{exoscale-impact-methodology}.
Scaleway allocates manufacturing impacts by reserved resource shares, so smaller reservations reduce this component, but the total also depends on CPU utilization after resizing~\cite{scaleway-instances-footprint}.
Azure's cost-based allocation rewards smaller VMs when they reduce normalized cost~\cite{microsoft-s3-2021}, and Oracle's spend-based method similarly rewards lower undiscounted charges~\cite{oracle-emissions-management}.
T~Cloud uses provisioned resources in its calculation, but its documentation does not establish how increased utilization after resizing affects the result~\cite{tcloud-sustainability-release,tcloud-sustainability-dashboard}.
Alibaba does not specify whether its resource quantities measure reserved capacity or actual activity~\cite{alibaba-cloud-carbon-footprint}, and OUTSCALE does not disclose the allocation weights needed to assess resizing~\cite{outscale-carbon-methodology}.

\parag{Region: Does using the same amount of electricity in a cleaner region lower reported emissions?}
Most providers publish location-based figures and use regional emission factors, including AWS~\cite{aws-ccft-model3}, GCP~\cite{schneider-mattia-2024}, Oracle~\cite{oracle-emissions-management}, OVHcloud~\cite{ovh-eit-v2}, Scaleway~\cite{scaleway-calculation-values} and Exoscale~\cite{exoscale-impact-methodology}.
The same electricity use is therefore attributed fewer emissions in a cleaner region.
Oracle provides this comparison through its power-based method, while its spend-based method is market-based only~\cite{oracle-emissions-management}.
IBM also documents location-based reporting, but currently lists that output as temporarily disabled~\cite{ibm-carbon-current}.
OUTSCALE's calculation likewise scales electricity emissions with the grid factor, although both currently covered regions use France's factor~\cite{outscale-carbon-methodology}.
Azure~\cite{azure-carbon-data} and T~Cloud~\cite{tcloud-sustainability-release} expose only market-based Scope~2 figures, while Alibaba includes green-electricity adjustments without documenting a separate location-based result~\cite{alibaba-cloud-carbon-footprint}.
Their regional totals may differ, but do not show how differences in grid carbon intensity affect reported emissions.

\parag{Time: Does running the same workload at cleaner times lower reported emissions?}
None of the reviewed methods documents a reduction in a customer's reported emissions from shifting work to cleaner hours.
GCP computes internal emissions hourly, but its published customer formula applies aggregate SKU and regional factors to customer usage without matching that usage to individual hours~\cite{schneider-mattia-2024}.
AWS allocates monthly service emissions in proportion to customer usage within that month~\cite{aws-ccft-model3}.
OVHcloud~\cite{ovh-eit-v2} and OUTSCALE~\cite{outscale-carbon-methodology} use yearly electricity emission factors.
IBM generally uses source factors updated yearly and applies them for up to a year~\cite{ibm-carbon-calc-v3}.
Scaleway~\cite{scaleway-calculation-values} and Exoscale~\cite{exoscale-impact-methodology} do not specify whether their electricity emission factors represent monthly or yearly averages.
Azure~\cite{azure-carbon-data}, T~Cloud~\cite{tcloud-sustainability-release} and Alibaba~\cite{alibaba-cloud-carbon-footprint} lack the separate location-based output needed for this comparison.
Oracle does not disclose the time resolution of its electricity factors or how customers' execution times enter the allocation~\cite{oracle-emissions-management}.

\endgroup

%% file: appendix-incentives.tex
\begingroup
\color{black}
\hypersetup{citecolor=black,linkcolor=black,urlcolor=black}
\section{Conditions for Allocation Incentives}
\label{app:operational-incentives}

The operational term of \RSCI decreases when a workload uses less energy or runs during cleaner hours or in cleaner regions.
By construction, total attributed emissions equal the provider's reported total.
The analysis below holds that report fixed to examine how allocation affects an individual workload's score, giving the incentive conditions discussed alongside Table~\ref{tab:properties}.

\parag{Setup}
Consider one reporting slice and suppress its index $s$.
Let $x\geq0$ be the workload's modeled operational emissions and $X>0$ the total across all workloads in the slice.
Hold reported emissions $C^{\downarrow k}\geq0$, scope fractions $\beta^k\in[0,1]$, functional output $R_i>0$, other workloads' operational emissions, and capacity shares fixed.
Write
\begin{equation}
A=\sum_k\beta^k C^{\downarrow k},\qquad
B=\sum_k(1-\beta^k)C^{\downarrow k},\qquad
b=\beta^{\mathrm{S2}},\qquad v=w^{\mathrm{cap}}_{i,s}\in[0,1].
\end{equation}
The energy and capacity residuals are $A-bX$ and $B-(1-b)X$, respectively.
With energy weights proportional to operational emissions, this workload's attributed emissions are
\begin{equation}
\begin{aligned}
F(x)&=x+\frac{x}{X}(A-bX)+v\bigl(B-(1-b)X\bigr)\\
    &=(1-b)x+A\frac{x}{X}+vB-v(1-b)X.
\end{aligned}
\end{equation}
Dividing by the fixed $R_i$ gives this slice's contribution to the workload's score.

\parag{Operational improvements}
Because other workloads' emissions are fixed, $dX/dx=1$, giving
\begin{equation}
\frac{dF}{dx}=(1-b)(1-v)+A\frac{X-x}{X^2}\geq0.
\label{eq:operational-incentive-derivative}
\end{equation}
\emph{Under these assumptions, reducing operational emissions cannot increase the reconciled score} while $X$ remains positive.
Both the residuals and the energy weights are recomputed after the change.
The derivative is strictly positive if either $b<1$ and $v<1$, or $A>0$ and other workloads have positive operational emissions.
In particular:
\begin{itemize}
\setlength{\itemsep}{1pt}
\item If $\beta^{\mathrm{S2}}<1$, a workload with less than the full capacity share receives a strictly lower score when its operational emissions decrease.
\item If $\beta^{\mathrm{S2}}=1$, positive reported Scope~2 emissions and positive operational emissions from other workloads suffice for a strict decrease. Energy allocations from other scopes can also make $A>0$.
\end{itemize}
These conditions cover energy savings and moving work to cleaner hours when operational emissions and energy weights use the time-varying carbon intensity, and capacity shares remain fixed.
The result extends across slices if the stated assumptions hold in each slice and operational emissions do not increase in any affected slice.
When work moves between slices, lower total operational emissions alone do not guarantee a lower score. The combined attributions must be compared.

\parag{Capacity allocation}
Now hold all operational emissions, reports, scope fractions, and output fixed, and vary the workload's capacity share $v$.
Its effect on the score is
\begin{equation}
\frac{\partial (F/R_i)}{\partial v}=\frac{B-(1-b)X}{R_i}=\frac{\Delta_s^{\mathrm{cap}}}{R_i}.
\end{equation}
\emph{A smaller capacity share lowers the score for a positive capacity residual, but raises it for a negative residual.}
If the capacity residual is zero, changing the capacity share has no effect on the score.
Setting every $\beta^k=1$ removes the capacity component.

\parag{Limits}
For a sole workload, $x=X$ and $v=1$, so $F=A+B=\sum_k C^{\downarrow k}$.
\emph{Its total attribution follows the provider's reported total.}
It therefore remains unchanged in the fixed-report analysis above.
These results concern attribution and do not establish physical emissions savings.
Actions that also change capacity shares, provider reports, scope fractions, other workloads' emissions, or functional output must be assessed with those changes included.
\endgroup